\documentclass[%
 reprint,
 amsmath,amssymb,
 aps,
]{revtex4-2}

\usepackage{graphicx}
\usepackage{dcolumn}
\usepackage{bm}

\usepackage[colorlinks=true, linkcolor=blue, citecolor=blue, urlcolor=blue]{hyperref}

\begin{document}

\title{A skin-like conformal sensor for real-time shape mapping}

\author{
	Kaiping~Yin$^{1\dagger}$,
	Sooik~Im$^{1\dagger}$,
	Chaorui~Qiu$^{1}$,
        Yun~Bai$^{1}$,
        Xiangyu~Lu$^{1}$,\\
        Chenhang~Li$^{1}$,
        Junjie~Yao$^{2,3}$,
        Xiaoyue~Ni$^{1,4,5\ast}$,
        \\
	\small$^{1}$Department of Mechanical Engineering and Materials Science, Duke University, Durham, NC 27708, USA.\\
	\small$^{2}$Department of Biomedical Engineering, Duke University, Durham, NC 27708, USA\\
        \small$^{3}$Department of Neurology, Duke University School of Medicine, Durham, NC 27710, USA.\\
        \small$^{4}$Department of Biostatistics and Bioinformatics, Duke University, Durham, NC 27710, USA.\\
	\small$^{5}$Department of Electrical and Computer Engineering, Duke University, Durham, NC 27708, USA.\\
	\small$^\ast$Corresponding author. Email: xiaoyue.ni@duke.edu\\
	\small$^\dagger$These authors contributed equally to this work.
}


\begin{abstract}
Reliable real-time 3D shape sensing is essential for robust control and interpretation of deformable systems during motion. Existing vision-based approaches require line-of-sight and complex instrumentation, limiting operation in occluded and space-constrained settings. Here, we introduce a scalable, skin-like sensor that reconstructs its continuous 3D deformation in real time from distributed strain measurements. The device embeds a 2D array of mirror-stacked, printed oxidized eutectic gallium–indium (o-EGaIn) strain gauges within an elastomeric film to measure off-neutral-axis strains. Combined with a mechanics-informed observation model and a fast optimization routine, the system estimates local curvature, elongation, offset, and orientation under concurrent stretching, bending, and indentation, enabling reconstruction of complex surfaces. A 5\,\texttimes\,5 array with a 12\,mm pitch achieves a mean surface reconstruction error of 0.62\,mm with $\sim$ 0.1\,s latency across all tested scenarios. When conforming to complex surfaces, the sensor provides fast 3D shape mapping of the underlying geometry. Demonstrations involving palm gesturing, finger indentation, and contact-induced balloon deformation highlight utility for epidermal motion tracking, haptic interaction, and intraoperative monitoring.
\end{abstract}

\maketitle


\section{\label{sec:intro}Introduction}

\noindent Soft and deformable systems increasingly underpin technologies in medicine, robotics, and engineering, creating a growing need for accurate shape information during operation.
The need is most acute in emerging conformal biomedical imaging and stimulation platforms, where the geometric configuration of distributed functional elements directly determines operational performance and fidelity (Fig.~\ref{fig:fig1}{a})~\cite{hu2023wearable, han2020catheter, zhu20203d, liu2024bioresorbable, darma2020real}. 
On-body wearables, such as stretchable and flexible 2D ultrasound array require accurate, real-time knowledge of transducer positions on the deformed body surface to support high-fidelity 3D beamforming and image reconstruction~\cite{hu2023wearable, zhou2024transcranial, zhang2026flexible}. Implantable and interventional devices face similar challenges; cardiac mapping balloons, for instance, require real-time registration of distributed electrodes as the balloon deforms against atrial tissue for accurate electroanatomic mapping and precise energy delivery~\cite{han2020catheter, kashyap2020multilayer}. Continuous, high-fidelity 3D tracking of the embedded elements across these settings is essential to maintain sensing fidelity, imaging resolution, and therapeutic accuracy. 

This need extends across a wider class of engineered soft systems. Soft robotic manipulators and continuum surgical instruments rely on proprioceptive shape estimation for stable grasping, occlusion-tolerant navigation, and dexterous teleoperation~\cite{lin2021arei, lilge2024state, glauser2019deformation, glauser2019interactive}. Human–machine interfaces and exosuits use it to infer posture, gesture, and motion intent~\cite{glauser2019deformation, glauser2019interactive}. Reconfigurable metasurfaces require shape readout to compensate wavefront distortion as they morph~\cite{li2021metasurfaces, li2025flexible}. Compliant aerospace structures depend on distributed shape information for fatigue and damage detection~\cite{gherlone2018shape, derkevorkian2013strain}. In each case, a thin conformal layer that reports its own 3D shape in real time would unlock spatial registration and closed-loop function that are currently out of reach.

Vision-based instrumentation currently dominates 3D shape acquisition of complex surfaces. Multi-view cameras~\cite{bai2022dynamically, feng2023real, vlasic2009dynamic}, photometric stereo~\cite{li2026biomimetic, yuan2017gelsight}, LiDAR~\cite{palieri2020locus}, and structured-light scanners~\cite{zhu20203d, zhu20183d} deliver high spatial resolution on accessible, well-lit surfaces.
These approaches falter, however, in occluded, contact-rich, or rapidly deforming settings. 
Embedded, lightweight, and flexible sensors have begun to address this gap by enabling in-situ proprioception.
For example, integrating functional materials (\textit{e.g.}, conductive~\cite{majidi2011non, medina2016resistor, pan2025shape, yamada2011stretchable}, optical~\cite{xu2017dual, xu2018stretchable, van2018soft,lun2019real,bai2020stretchable}, dielectric~\cite{glauser2019deformation, navarro2020model}, piezoelectric~\cite{rendl2014flexsense}, etc.) into a deformable surface transduces local axial strain into electrical signals. Marrying rigid magnetic flux sensors~\cite{magnet2019, lilge2022continuum, xiang2026portable} or inertial measurement units (IMU)~\cite{mittendorfer20123d,zhou20233d,shah2023stretchable} with a soft substrate provides local orientation. 

Despite these advances, real-time, accurate, and robust 3D surface reconstruction from discrete, sparse sensor nodes remains a central challenge, characterized by the competing trade-offs between hardware complexity, computational tractability, and model fidelity. 
Most physics-based methods preserve tractability by assuming a non-stretchable backbone or array~\cite{al2020improved, medina2016resistor, zhou20233d,li2025flexible}. This assumption fails for non-developable surfaces, which inherently undergo coupled bending, stretching, and compression~\cite{modes2011gaussian, ha2024stretchable, liu2023conformability}. Decoupling multi-modal strain typically demands heterogeneous sensing modalities~\cite{kim2020heterogeneous, shah2023stretchable} or heterostructured sensing configurations~\cite{white2017multi, bai2020stretchable, leber2020soft}, with footprint, wiring, and crosstalk costs that scale poorly to dense 2D arrays.
Hardware advances alone also face an algorithmic bottleneck: Propagating sparse local measurements into global geometry can compound errors and produce pronounced shape deviations~\cite{xu2017dual, lilge2022continuum, zhou20233d}. Global constraints can mitigate error accumulation, but no existing approach embeds them in a formulation that is at once accurate, well-posed, and efficient for real-time inference without supplemental external sensing~\cite{an2024shape, liu2025model, lu2023robust}.

Data-driven approaches, particularly deep learning, can map sparse sensor signals to 3D surface geometry with simple hardware and offer rapid post-training inference~\cite{lun2019real, hu2023stretchable, scharff2021sensing, van2018soft, ha2022shape}. 
Performance, however, often degrades when the measurements do not uniquely determine the underlying deformation, an observability gap rooted in missing multimodal-strain information. Networks then regress toward shapes that dominate the training distribution, producing plausible but incorrect reconstructions whose limited generalization and weak interpretability are practical liabilities in applications demanding reliability and traceable failure modes~\cite{hao2023online, wang2022generalizing}.

Real-time, proprioceptive mapping of nontrivial 3D shapes requires two advances: 1) scalable, low-profile hardware that supports dense, precisely coordinated sensor networks over highly curved surfaces, and 2) an efficient computational framework that fuses distributed local measurements into accurate, robust global shape reconstructions across diverse deformation modes. 
To address the unmet need, we present a soft, skin-like conformal sensor that maps complex 3D shapes in real time. This system combines multilayer, stretchable strain-gauge arrays with a fast reconstruction pipeline that remains accurate under concurrent bending, stretching, and indentation. Key contributions include: (1) laser-patterned liquid-metal-based strain gauge with materials, geometry, and layout optimized for high performance and scalable fabrication; (2) a low-profile, multilayer architecture that supports dense, easily routed arrays of individually addressable sensors; (3) a mechanical model that infers local curvature and elongation from dual-layer strain measurements; and (4) a physics-based optimization algorithm that exploits both local deformation and global geometric constraints to efficiently and accurately reconstruct continuous 3D surfaces. The soft, thin-film form factor allows seamless conformation to complex, dynamically changing surfaces. The architecture readily functions as an encapsulation layer compatible with integration into broader soft device platforms, preserving the performance of systems that require precise spatiotemporal registration of distributed functional elements. Such a system provides unprecedented in-situ access to shape information of deformable engineering systems operating in real-world environments where clear, stable visual access is not feasible. It establishes a route to real-time proprioceptive shape mapping framework and provides unique opportunities to transform how soft systems perceive, adapt, and function across diverse mechanical environments.

\section{\label{sec:css}The conformal shape sensor}

\begin{figure*}
    \centering
    \includegraphics[width=0.9\textwidth]{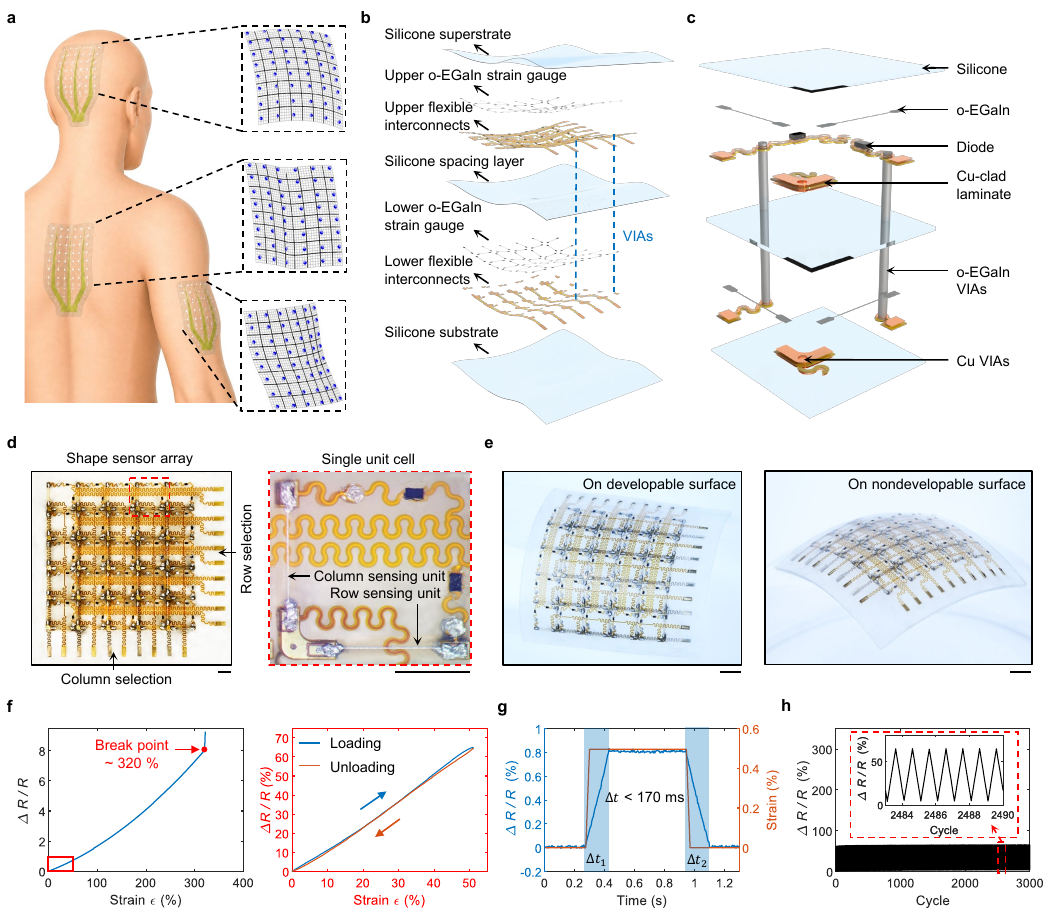}
    \caption{\textbf{Skin-like conformal shape sensor based on mirror-stacked, individually addressable soft strain gauges.}
        {\bf a}, Conceptual use case in which the shape sensor integrates with medical devices bearing distributed functional elements (e.g., transducers or electrodes), maps local anatomy, and proprioceptively reconstructs the 3D coordinates of each element.
        {\bf b}, Exploded schematic of the device architecture. 
        {\bf c}, Enlarged exploded view of one unit cell in the sensor grid. 
        {\bf d}, Optical images of the sensor array (top view) with a magnified view of a single unit cell (red dashed box). Scale bars, 5\,mm. 
        {\bf e}, Photographs of the as-fabricated device conforming to a developable surface and a non-developable surface. Scale bars, 10\,mm. 
        {\bf f}, Relative resistance change ($\Delta R/R$) of the o-EGaIn strain gauge as a function of uniaxial tensile strain ($\epsilon$), showing stretchability to $\sim320$\% without failure (left) and  a linear, reversible range ($R^2>0.998$) over 50\% (right).
        {\bf g}, Transient response to a 0.5\% step strain (strain rate, 17\%\,s$^{-1}$). {\bf h}, Cyclic stretch-release stability under 50\% uniaxial tensile strain (strain rate, 2.5\%\,s$^{-1}$). }

    \label{fig:fig1}
\end{figure*}

Fig.~\ref{fig:fig1}{b} presents an exploded view of the shape sensor. Fig.~\ref{fig:fig1}{c} shows a unit cell containing two orthogonal sensing units. Each sensing unit comprises a mirror-stacked pair of dog-bone-shaped o-EGaIn strain gauges arranged symmetrically about a 400\,\textmu m-thick spacing layer of silicone elastomer (Ecoflex 00-30, Smooth-On). Stirring mix 75 wt.\% Ga and 25 wt.\% In forms the paste-like o-EGaIn ink that wets on silicone with an adhesion force of 0.8\,mN~\cite{zhang2022controlling, kaneko2023printed, ma2023shaping}. Stencil printing defines the gauge layout on silicone, followed by laser hatching to narrow the conductive channel and increase baseline resistance. This process yields a nominal gauge geometry of 1.0\,cm\,\texttimes\,75.0\,\textmu m\,\texttimes\,7.5\,\textmu m, used throughout the study.
Two layers of flexible interconnects, patterned from a copper (Cu)-clad laminate (35-\textmu m-thick top and bottom Cu layer, 25-\textmu m-thick middle PI layer; AP9111R, DuPont Pyralux), 
route the gauge signals. 
A 200\,\textmu m-thick silicone substrate and a 350\,\textmu m-thick superstrate encapsulate the stack.
An o-EGaIn-filled vertical interconnect access (VIA; 400\,$\mu$m in diameter) connects the upper and lower gauges within each sensing unit. A diode switch pairs with both gauges, enabling independent addressing and simultaneous readout of the two channels~\cite{prutchi1993IDM}. This layout organizes the sensing units into a row-column addressable array and uses time-division multiplexing (TDM) to sequentially access each unit through edge-distributed electrodes. An $N$\,\texttimes\,$M$ array contains $2NM$ sensing units, requiring only $2(N+M)$ data acquisition (DAQ) channels to read. A scan signal at frequency $F_{sc}$\,Hz yields an effective per-unit sampling rate of 0.5\,$F_{sc} / \text{min}(N, M)$\,Hz per sensing unit.
Fig.~\ref{fig:fig1}{d} shows an as-fabricated 5\,\texttimes\,5 shape sensor array, with an enlarged view of a representative unit cell. The array has a pitch size of $a$\,=\,12\,mm and integrates 100 individually addressable strain gauges over an active area of approximately 40\,cm$^2$. The overall device exhibits an effective Young's modulus of 562\,kPa and an estimated bending stiffness of $\sim$ 0.083\,N$\cdot$mm, both within the commonly cited comfort range of skin-mounted wearable devices (420-850\,kPa and $<\sim$ 0.5\,N$\cdot$mm, respectively) \cite{lee2020mechano, huang2018three, hu2023wearable}. The skin-like mechanics enable conformal integration with both developable and non-developable surfaces, as demonstrated in Fig.~\ref{fig:fig1}{e}.

The adopted o-EGaIn gauge design features a combination of high stretchability, high sensitivity, low hysteresis, and high repeatability, which are essential for high-fidelity dynamic shape extraction ~\cite{kong2020oxide, kaneko2023printed, xu2025precision}. Fig.~\ref{fig:fig1}{f} presents the fractional change in resistance of a single gauge channel, $\Delta R/R$, as a function of the uniaxial strain $\epsilon$ at a strain rate of 0.7\%\,s$^{-1}$. The gauge tolerates extreme deformation; it sustains up to $\sim$320\% tensile strain without failure, and maintains a linear, reversible electrical response ($R^2 > 0.998$) over 50\% strain with a gauge factor $GF$ of 1.26. By contrast, the flexible interconnects exhibit negligible deformation-induced resistance change and therefore do not measurably bias the strain readout. Within the linear regime, the signal exhibits low hysteresis error ($\gamma_H$ = 3.52\%)~\cite{huang2023high}. 
The fabricated gauges achieve 100\% yield and show low channel-to-channel variability, with $R$ = 6.55\,$\pm$\,0.72\,$\Omega$, $GF$ = 1.24\,$\pm$\,0.19, and $\gamma_H$ = 4.21\,$\pm$\,0.70\% across 100 channels. Representative resistance noise exhibits a standard deviation (s.d.) of $\sigma_n$ = $\sim$\,0.007\%, corresponding to a 3$\sigma_n$ strain resolution of $\sim$\,0.017\%.
Fig.~\ref{fig:fig1}{g} shows sensor response to a rapid 0.5\% step stretching at a strain rate of $\sim$\,17\%\,s$^{-1}$. Response time $\Delta t_1$ and recovery time $\Delta t_2$ are both below 170\,ms. Resistance response during stress relaxation under constant-strain holding shows a relaxation fraction of 3.5\% at 20\% strain. Fig.~\ref{fig:fig1}{h} presents cyclic testing of a single channel over 3,000 cycles at a 50\% strain amplitude and a strain rate of 2.5\%\,s$^{-1}$. Both $R$ and $GF$ remain stable throughout the cycles, with a maximum deviation of 4.2\% in $R$ and 5.6\% in $GF$.

\section{\label{sec:workingprinciple}Working principles of mode decoupling and curve reconstruction}
\begin{figure*}
    \centering
    \includegraphics[width=0.85\textwidth]{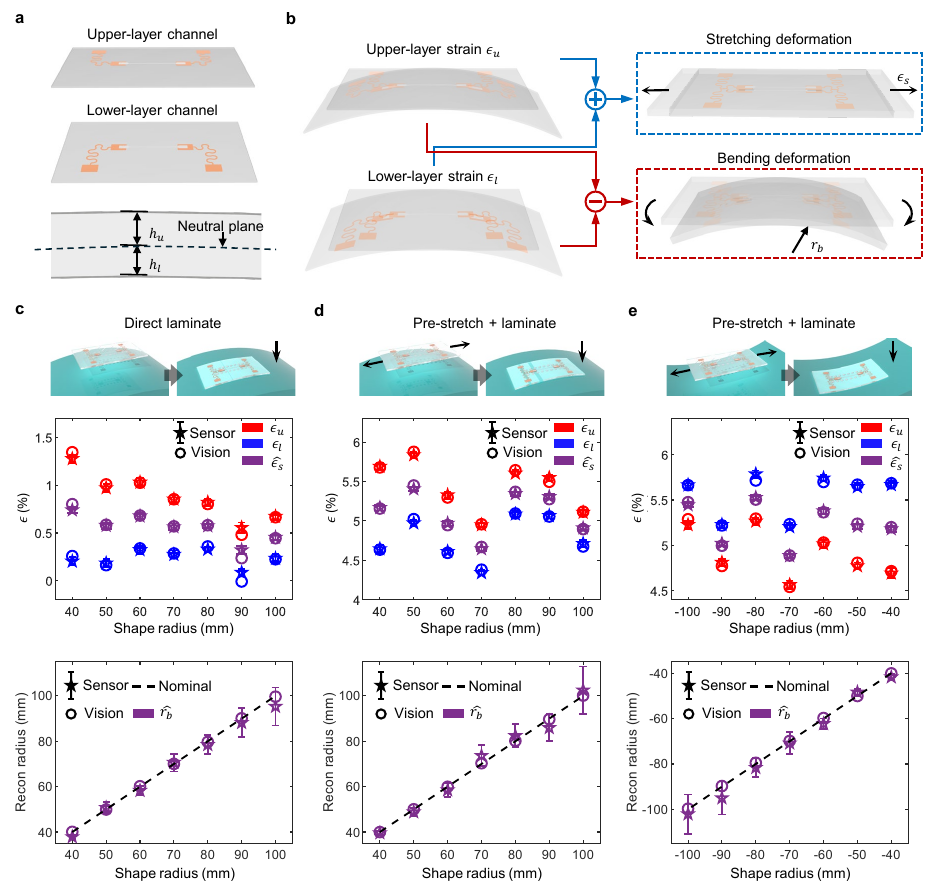}
    \caption{\textbf{Working principles of the sensing unit for decoupling stretching and bending.}
        {\bf a}, Schematic of the sensing unit showing an exploded side view of the mirror-stacked o-EGaIn strain gauges and a thickness-view indicating the neutral plane. 
        {\bf b}, Two-layer readout for mode separation: the common-mode strain (layer average) quantifies stretching, whereas the differential-mode strain (layer difference) quantifies bending. 
        {\bf c\,--\,e}, Constant-curvature mold tests (nominal radius $r_b$\,= 40\,--\,100\,mm, 10\,mm interval) under three mounting conditions: direct lamination onto a convex mold ({\bf c}), pre-stretch followed by lamination onto a convex mold ({\bf d}), and pre-stretch followed by lamination onto a concave mold ({\bf e}). In {\bf c\,--\,e}, the panels show, from top to bottom, schematics of the mounting procedure, upper- and lower-layer strain measurements with the corresponding stretching ratio estimation $\hat{\epsilon_s}$, compared with the vision-based reference across mold radii, and the sensor-reconstructed radii compared with the vision-derived and nominal radii. For each radius, the sensor data shows one representative measurement, with the error bars indicating the standard deviation (s.d.) of the difference between the sensor and vision measurements ($n$\,=\,20 repeated measurements on the same device).}
    \label{fig:fig2} 
\end{figure*}

A central challenge in shape reconstruction with stretchable strain gauges arises from the concurrent contributions of bending and stretching to the sensor output. This work advances a mechanical model that exploits the mirror-stacked gauge design to decouple bending from stretching \cite{white2017multi}. Fig.~\ref{fig:fig2}{a} illustrates the structure of a single sensing unit with upper-layer (subscript $u$) and lower-layer (subscript $l$) gauges, where the two sensor layers lie at thickness-direction offset $h_u$ and $h_l$ from the neutral plane, respectively. Under the Kirchhoff–Love kinematic assumption, the normal strain at a distance $h$ from the neutral plane decomposes into a stretching component $\epsilon_s$ and a bending component $\epsilon_b = h / r_b$, where $r_b$ is the signed radius of curvature. The measured upper- and lower-layer strains, $\epsilon_u$ and $\epsilon_l$, therefore satisfy:
\begin{equation}
\begin{split}
    \epsilon_u = (1+\epsilon_s) (1 + \frac{h_u}{r_b})-1, \\
    \epsilon_l = (1+\epsilon_s) (1 - \frac{h_l}{r_b})-1.
\end{split}
\label{eq:init_governing}
\end{equation}
Here, $h_u$ and $h_l$ denote the absolute thickness offsets of the upper and lower gauges from the neutral plane (Fig.~\ref{fig:fig2}{a}), and $r_b >$ 0 for convex curvature (upper layer in tension) and $r_b <$ 0 for concave curvature. With this convention, $\epsilon_u$ and $\epsilon_l$ capture both the sign and magnitude of bending coupled with stretching. The average of the two strains (common mode) captures stretching, whereas their difference (differential mode) captures bending (Fig.~\ref{fig:fig2}{b}). Under a circular-arc assumption, the estimated radius $\hat{r_b}$ and stretching ratio $\hat{\epsilon_s}$ fully characterize the deformation and satisfy:
\begin{equation}
    \hat{r_b}=\frac{(1+\epsilon_l) \cdot h_u + (1+\epsilon_u)\cdot h_l}{\epsilon_u - \epsilon_l},
    \label{eq:calc_radius}
\end{equation}
\begin{equation}
    \hat{\epsilon_s} = \frac{\epsilon_uh_l + \epsilon_lh_u}{h_u + h_l},
    \label{eq:calc_eps}
\end{equation}
where $h_u$ and $h_l$ are design parameters; $\epsilon_u$ and $\epsilon_l$ are measured quantities. Two error sources affect the estimation accuracy of $\hat{r_b}$ and $\hat{\epsilon_s}$: (i) measurement error in $\epsilon_u$ and $\epsilon_l$, and (ii) uncertainty in $h_u$ and $h_l$. 
A parametric sweep analysis shows that increasing the total thickness, $h_u + h_l$, reduces the sensitivity of $\hat{r_b}$ to both error sources. However, this benefit plateaus beyond $\text{0.4\,mm}$, whereas the bending stiffness continues to increase sharply. In contrast, the sensitivity of $\hat{\epsilon_s}$ remains nearly unchanged with $h_u + h_l$. Together, these results support choosing $h_u+h_l$ = 0.4\,mm, which keeps the maximum theoretical error ratios in $\hat{r_b}$ and $\hat{\epsilon_s}$ below 10\%.

Fig.~\ref{fig:fig2}{c--e} characterize the performance of a sensing unit laminated on 3D-printed cylindrical molds with radii from 40\,mm to 100\,mm in 10\,mm increments. Three mounting configurations impose distinct combinations of bending and stretching: 1) direct lamination onto a convex mold; 2) lamination onto a convex mold after $\sim$\,5\% pre-stretching; and 3) lamination onto a concave mold after the same pre-stretch. Stereo-vision tracking of o-EGaIn fiducial markers co-printed with the upper- and lower-layer gauges 
provides a ground-truth strain measurement. Across all configurations, strains measured from both channels closely match the vision-based reference, with a maximum deviation of 0.099\%. The corresponding estimates, $\hat{r_b}$ and $\hat{\epsilon_s}$, show maximum errors of 16.4\,mm and 0.074\%. 
At a given strain or thickness uncertainty, errors in $\hat{r_b}$ and $\hat{\epsilon_s}$ vary with the nominal values $r_b$ and $\epsilon_s$. Their propagation to the endpoint coordinates of the fitted circular-arc segment, however, only produces a maximum endpoint deviation ratio of 1.8\% across all tested cases.

\section{\label{sec:3drecon}3D surface reconstruction}
\begin{figure*}
    \centering
    \includegraphics[width=1\textwidth]{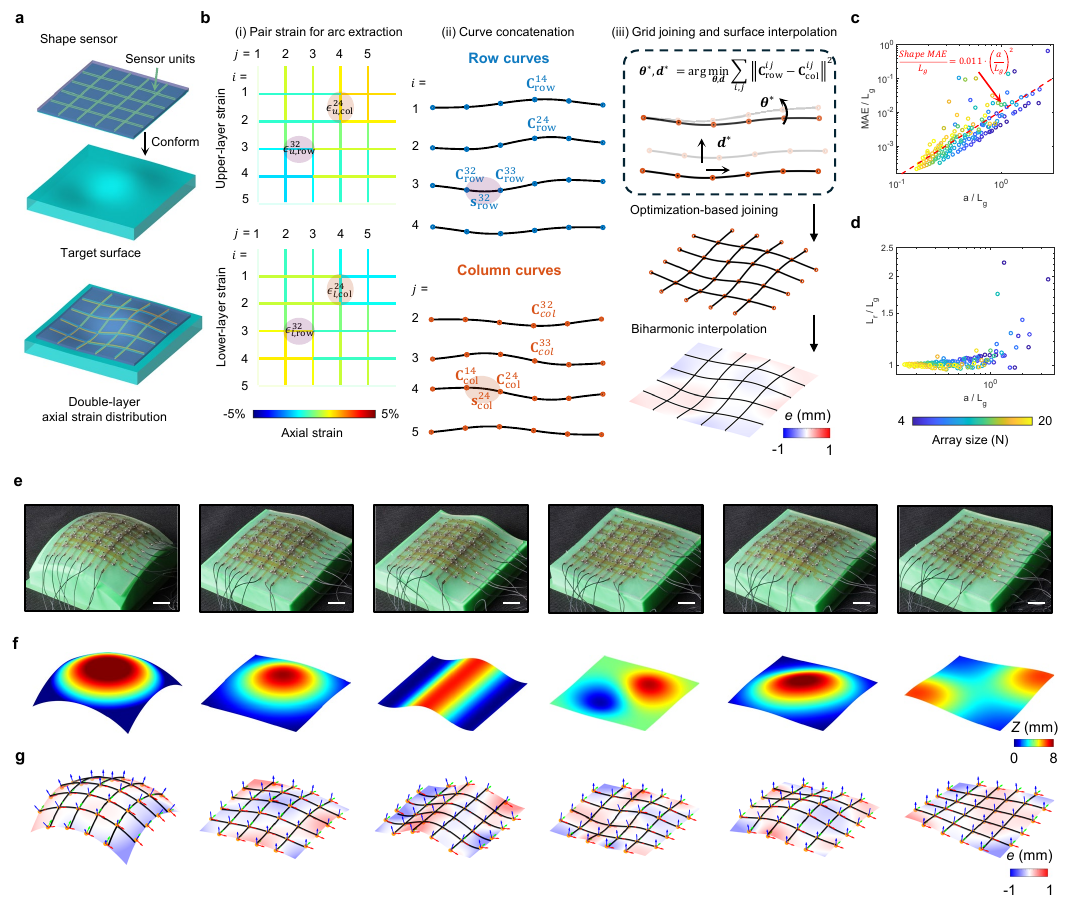}
    \caption{\textbf{3D reconstruction process and conformal shape mapping of rigid surfaces.}
        {\bf a}, Finite element simulation setup of a 5\,\texttimes\,5 shape sensor conforming to a rigid surface. {\bf b}, Reconstruction workflow illustrated using the simulated strain fields in {\bf a}: {(i)} pair upper- and lower-layer strain signals to define local circular arcs, {(ii)} concatenate neighbouring arcs to form row and column curves, and {(iii)} join the curve families into a coherent grid and interpolate it to generate a surface point cloud. 
        {\bf c}, Simulated reconstruction error normalized by deformation feature size (MAE$/L_g$) and 
        {\bf d}, feature reproduction fidelity ($L_r/L_g$) as a function of sensor sparsity ($a/L_g$) for arrays of varying size ($N$); the red dashed line indicates the fitted scaling. 
        {\bf e}, Optical images of the sensor mounted on six different 3D-printed molds. Scale bars, 10\,mm. 
        {\bf f}, Ground-truth parametric geometries of the six molds. 
        {\bf g}, Reconstructed surfaces with sensor-grid lines overlaid and reconstruction error shown as color contours. Orange markers denote ground-truth grid intersections, and local coordinate frames indicate the corresponding ground-truth orientations.}
    \label{fig:fig3} 
\end{figure*}

Reconstructing a continuous surface using an $N$\,\texttimes\,$M$ shape sensor array requires fusing the local geometric estimates ($\hat r_b$ and $\hat\epsilon_s$) from all the $2NM$ sensing units.
Fig.~\ref{fig:fig3}{a} shows the finite element simulation setup of a 5\,\texttimes\,5 device laminated onto a rigid, bivariate Gaussian surface.
Fig.~\ref{fig:fig3}{b} shows the simulated upper ($u$) and lower ($l$) strain fields in row and column directions, $\boldsymbol{\epsilon}_{u} = \{\epsilon^{ij}_{u, \text{row}}, \epsilon^{ij}_{u, \text{col}}\}$ and $\boldsymbol{\epsilon}_{l} = \{{\epsilon^{ij}_{l,\text{row}}, \epsilon^{ij}_{l,\text{col}}}\}$, where $(i, j)$ are row- and column-index of the unit cell. Using \eqref{eq:calc_radius} and \eqref{eq:calc_eps}, each upper- and lower-strain pair defines a circular-arc segment, represented by $\mathbf{s}_\text{row}^{ij}=\{\epsilon^{ij}_{s,\text{row}}$, $r^{ij}_{b,\text{row}}\}$ and $\mathbf{s}_\text{col}^{ij} = \{\epsilon^{ij}_{s,\text{col}}$, $r^{ij}_{b,\text{col}}\}$. Assuming a continuous curvature gradient, concatenating neighboring segments along two principal directions yields two families of smooth curves, with junction coordinates $\mathbf{C}^{ij}_\text{row}$ and $\mathbf{C}^{ij}_\text{col}$ (Fig.~\ref{fig:fig3}{b}). 
The geometric consistency requires row and column curves to share a common node at each intersection. An intersection-joining procedure enforces this constraint by updating the absolute orientations $\boldsymbol{\theta}=\{\theta_\text{row}^i, \theta_\text{col}^j\}$ and translations $\boldsymbol{d}=\{\boldsymbol{d}_\text{row}^i, \boldsymbol{d}_\text{col}^j\}$ of the two curve families, minimizing the summed Euclidean distance between their predicted junction points:
\begin{equation}
\label{eq:curve_pose_opt}
\begin{aligned}
\{\boldsymbol{\theta}^*, \boldsymbol{d}^*\}
&= \arg\min_{\boldsymbol{\theta},\,\boldsymbol{d}}
\sum_{(i,j)}
\left\|
\mathbf{C}_\text{row}^{\,ij}(\boldsymbol{\theta},\boldsymbol{d})
-
\mathbf{C}_\text{col}^{\,ij}(\boldsymbol{\theta},\boldsymbol{d})
\right\|_2^{2}.
\\
\end{aligned}
\end{equation}
Solving the optimization problem yields $\boldsymbol{\theta}^*$ and $\boldsymbol{d}^*$ (Fig.~\ref{fig:fig3}{b}). The formulation reduces the problem to a constrained quadratic programming, enabling an efficient solution. After curve joining, 3D interpolation converts the grid into a semi-continuous surface point cloud at 400\,pts$\cdot a^{-2}$. 

Representative reconstruction of a 5\,\texttimes\,5 shape sensor from the geometry-sweep simulation yields an average mean absolute error (MAE) of $\sim$ 0.35\,mm and a feature reproduction fidelity ($L_r/L_g$) of 1.30. Here, $L_r$ and $L_g$ denote the feature sizes of the reconstructed and ground-truth shapes, respectively. In each case, the feature size ($L$) is the diameter of the largest circle enclosed by the planar -6\,dB contour around the local peak. Fig.~\ref{fig:fig3}{c, d} show that, regardless of the $N$\texttimes$N$ sensor array size ($N\in$ [4, 20]), both the normalized MAE (MAE$/L_g$) and $L_r/L_g$ degrade as the relative sensor sparsity, quantified by the normalized pitch size $a/L_g$, increases. When $a/L_g < 1$, MAE$/L_g$ remains generally below 0.040 while the feature-size deviation, $\|L_r/L_g-1\|$ stays below 10\%. 

Fig.~\ref{fig:fig3}{e} presents optical images of a 5\,\texttimes\,5 shape sensor mounted on six 3D-printed rigid surfaces. The parametric geometries include both developable and non-developable forms, with a typical feature size $L_g$\,=\,15\,mm ($a/L_g$\,=\,0.8) (Fig.~\ref{fig:fig3}{f}). Fig.~\ref{fig:fig3}{g} visualizes the proprioceptively reconstructed surface with the sensor grid, overlaid with error contours, ground-truth intersection coordinates and orientations. Across six surfaces, MAE peaks at 0.85\,mm (MAE/$L_g$\,=\,0.071), with an average MAE of $\sim$\,0.51\,mm (MAE$/L_g$ = 0.039), exceeding the Monte Carlo–derived MAE of $\sim$\,0.19\,mm (MAE$/L_g$\,=\,0.013) obtained under uncertainties at the strain resolution level of 0.017\%. Experimental deviations likely stem from through-thickness shear that weakly violates the Euler–Bernoulli assumption, together with electrical noise and related readout nonidealities \cite{ochsner2021classical, boujamaa2008rejection}. The reconstruction runtime for a 5\,\texttimes\,5 system is $\sim$\,0.08\,s and exhibits a subquadratic scaling with the array size $N$ for $N$\,\texttimes\,$N$ shape sensor arrays.

\section{\label{sec:dynamic}Real-time shape mapping of dynamically-morphing surfaces}
\begin{figure*}
    \centering
    \includegraphics[width=1\textwidth]{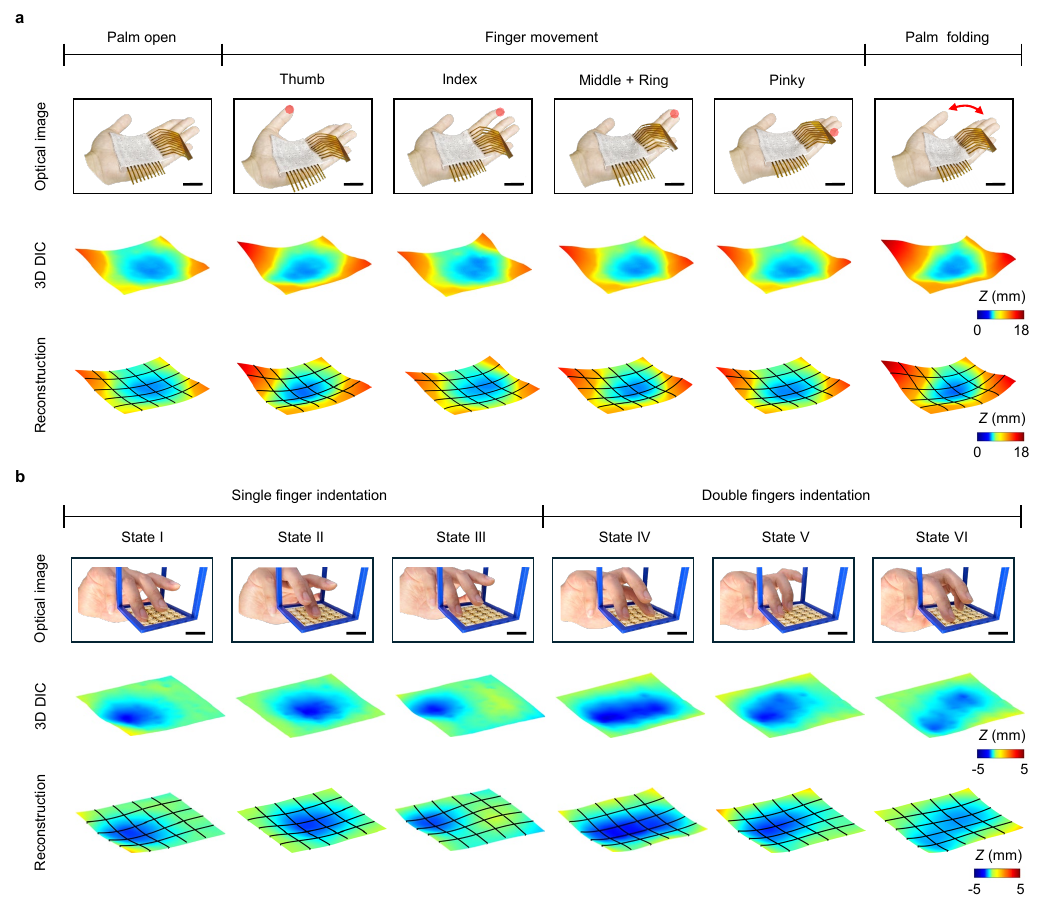}
    \caption{\textbf{Real-time shape mapping for epidermal motion tracking and haptic interaction.}
        {\bf a}, A 5\,\texttimes\,5 array tracks dynamic palm deformations during representative gestures. 
        {\bf b}, A 5 × 5 array tracks localized finger indentation with all four edges fixed. Representative frames show, from top to bottom, optical images, 3D-DIC ground-truth contours, and shape-sensor reconstructions. Scale bars, 20\,mm.
        }
    \label{fig:fig4} 
\end{figure*}

The skin-like shape sensor enables real-time shape mapping of deformable surfaces under dynamic loading, with the 5\,\texttimes\,5 sensor outputting reconstructions at $\sim$10\,Hz. Fig.~\ref{fig:fig4}{a} demonstrates epidermal motion tracking by mounting the sensor on a palm and capturing bending-dominated shape morphing across six representative gestures. A top-view stereo-vision setup with speckle patterns on the sensor surface provides 3D digital image correlation (3D-DIC) ground truth. Across all frames, the deformation reaches a maximum relative sparsity ($a/L_g$) of $\sim$\,0.61. Under this condition, reconstructions agree with 3D-DIC with a maximum MAE of 0.81\,mm (MAE$/L_g$ = 0.040) and a maximum $\|L_r/L_g-1\|$ of 4\%. 

Fig.~\ref{fig:fig4}{b} illustrates a scheme where finger indentation produces substantial stretching of the suspended membrane with fixed boundaries. Single- and double-finger trials generate representative, non-redundant deformation modes, while below-view 3D-DIC captures the ground truth. $a/L_g$ peaks at 0.86 across all frames. Sensor reconstructions match the ground truth, with a maximum MAE of 0.63\,mm (MAE$/L_g$ = 0.045) and a maximum $\|L_r/L_g-1\|$ of 9\%. 

\section{\label{sec:balloon}Proprioceptive shape sensing in a balloon surrogate during inflation and contact}

\begin{figure*}
    \centering
    \includegraphics[width=1\textwidth]{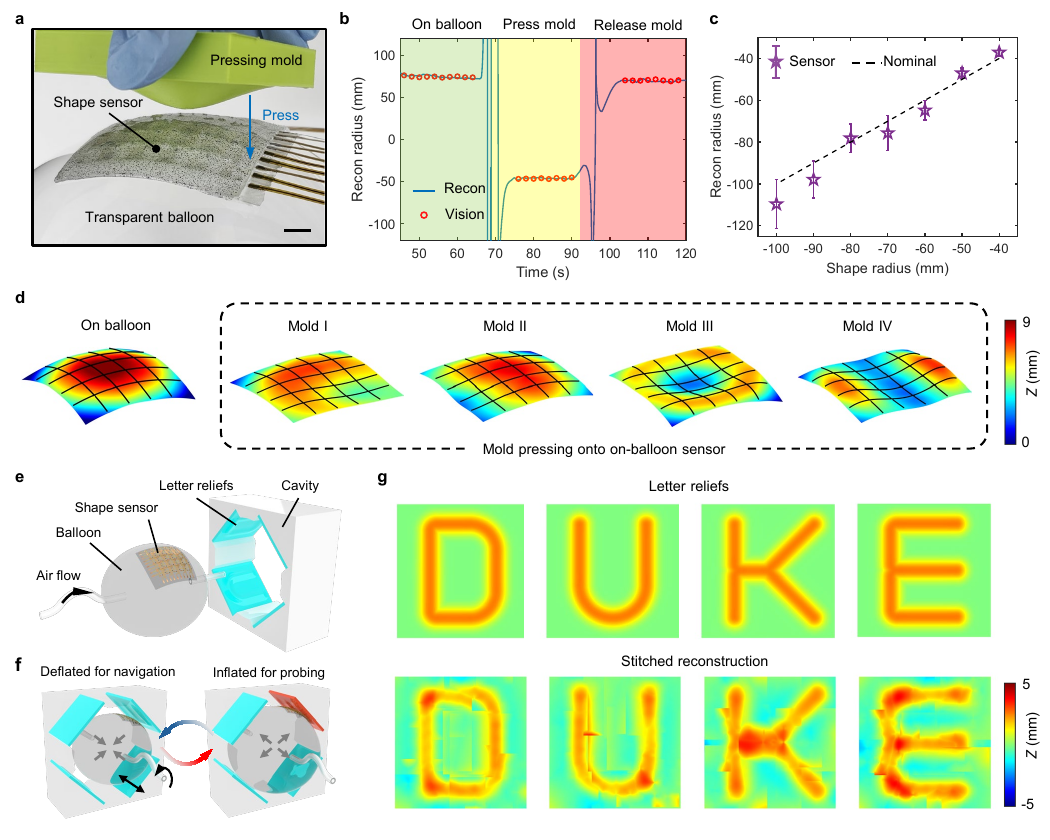}
    \caption{\textbf{Sensor-instrumented balloon demonstrating proprioceptive shape mapping.}
        {\bf a}, Optical image of a rigid, dome-shaped mold (radius, 50\,mm) indenting an inflated balloon with integrated 5\,\texttimes\,5 shape sensor. Scale bar, 10 mm. 
        {\bf b}, Radius reconstructed from a representative sensing unit over time before, during, and after indentation, with 3D-DIC reference. 
        {\bf c}, Reconstructed radius versus nominal mold radius for indentation with dome-shaped molds of varying radii. Data are mean $\pm$ s.d. ($n$\,=\,20 repeated measurements on the same device). 
        {\bf d}, Reconstructed surfaces during indentation using four molds of different shapes (dashed box; mold I\,--\,IV). 
        {\bf e}, Schematic of a sensor-instrumented balloon probe for intracavitary shape mapping in a rigid cavity: the deflated state enables navigation by displacement and rotation, whereas inflation establishes contact for indentation-based shape probing. 
        {\bf f}, Probing workflow for a cavity with inner wall patterned with four letter-shaped reliefs. 
        {\bf g}, Letter-shaped reliefs (top) and stitched shape reconstructions assembled from multiple probe measurements for each mold (bottom).}
    \label{fig:fig5} 
\end{figure*}

The conformal shape sensor provides proprioceptive sensing when integrated with soft structures. In clinical catheter–balloon platforms, analogous shape information can support cardiac potential mapping and electrogram acquisition, and help quantify tissue contact during ablation~\cite{han2020catheter, xu20143d, cluitmans2017vivo}. A controlled benchtop balloon experiment presents this capability. Fig.~\ref{fig:fig5}{a} shows the sensor mounted on a pre-inflated polyurethane balloon (0.15\,mm thickness) with an initial radius of 80\,mm. Double-sided adhesive tape (2477P, 3M; 5,mm width) secures the sensor perimeter, while the non-bonded contact over the central body minimizes parasitic shear loading. A spherical-dome mold (radius $\sim$50\,mm, relief height 7\,mm) indents the sensor area on the balloon and drives local conformation to the imposed geometry. Fig.~\ref{fig:fig5}{b} plots the arc radius estimation ($\hat{r_b}$) measured from one representative unit inside the sensor array, which reversibly switches between convex and concave curvature during press and release. Across molds of varying radii, reconstructed curvature follows the applied curvature with a maximum deviation of 20.9 mm (20.9\% at mold radius of 100 mm; Fig.~\ref{fig:fig5}{c}). Four additional 3D-printed molds with more complex geometries (with a minimum $L_g$ of 15\,mm) further challenge the reconstruction (Fig.~\ref{fig:fig5}{d}). 3D-DIC ground truth from the bottom view indicates a maximum MAE of 0.87\,mm (MAE$/L_g$ = 0.058) and a $\|L_r/L_g-1\|$ peaks at 5\%. The sensor-instrumented balloon proprioceptively tracks shape during inflation, deflation, and contact while maintaining compliant balloon deformation.

The same balloon platform also enables tactile mapping when visual access is limited. In an intracavitary demonstration, the deflated balloon enters a rigid cavity and then inflates to press the sensor area against an inner wall patterned with 3D-printed letter reliefs (`D', `U', `K' and `E') (Fig.~\ref{fig:fig5}{e}). Each letter relief spans 120\,mm\,\texttimes\,120\,mm, with a minimum feature size of 15\,mm and a maximum relief height of 4\,mm. Manual rotation and translation bring successive wall regions into contact with the sensor (Fig.~\ref{fig:fig5}{f}). Multiple contact events stich together to reconstruct the full letter geometry. Fig.~\ref{fig:fig5}{g} shows the reconstructed shapes of all four letters after subtracting the baseline (pre-contact) surface, demonstrating the system's ability to recover complex geometries through distributed, tactile exploration.

\section{\label{sec:Conclusions}Conclusions}
This work establishes a scalable, skin-like 3D shape sensor that reconstructs continuous surface geometry from distributed measurements of its own deformation in real time. Mirror-stacked, stretchable strain gauges provide off-neutral-axis readout that decouples axial stretching from bending. A physics-based reconstruction pipeline converts local geometric estimates into global 3D surface coordinates. Across all demonstrations, a 5\,\texttimes\,5 array achieves an average MAE of 0.62\,mm at $\sim$10\,fps. This hardware–algorithm combination achieves a practical balance between reconstruction fidelity and computational efficiency, making it uniquely suited for accurate real-time 3D shape mapping.

The thin, compliant sensor conforms to high-curvature, dynamic surfaces and operates without line-of-sight, enabling proprioceptive shape readout and real-time registration of distributed elements on deformable substrates. Higher spatial resolution and reconstruction accuracy motivate denser arrays. The system framework reported here provides a practical path to larger array sizes. This scalability arises from high-throughput fabrication methods (printing, laser cutting, and multi-layer integration), as well as multiplexed readout schemes and efficient reconstruction algorithms that keep wiring and computational complexity manageable as array size grows. Miniaturization, however, entails shorter gauges with lower base resistance, which makes the readout more susceptible to parasitic series resistance and thermal drift \cite{liu2024three}. Achieving finer grids therefore requires further advancement in gauge materials, sensor architectures, and readout strategies that preserve both sensitivity and mechanical performance in compact formats.

System-level integration and mobility offer further opportunities. On-sensor multiplexing and signal conditioning, combined with wireless power and communication, can eliminate tethering and improve performance during ambulatory or remote operation \cite{chen2026noise}. Coupling proprioceptive shape feedback with embedded shape actuation could further enable closed-loop control of geometry and functional response in soft robots and wearables \cite{lilge2024state, ha2022shape}. Multimodal integration (for example, normal/shear pressure or temperature) can extend the platform to further advance contact-mechanics inference for haptic interaction and manipulation \cite{li2026biomimetic}. Together, these opportunities point to soft electronic systems that merge distributed perception with adaptive physical function, thereby establishing a foundation for intelligent, interactive and autonomous biointegrated and robotic platforms.

\begin{acknowledgments}
\noindent The authors gratefully acknowledge financial support from the Beyond the Horizon Initiative of the Pratt School of Engineering at Duke University; the American Heart Association Collaborative Sciences Award (25CSA1417550); and the National Institute of Biomedical Imaging and Bioengineering of the National Institutes of Health under Awards R21EB034973 and R01EB037095. The content is solely the responsibility of the authors and does not necessarily represent the official views of the funding agencies. This work was performed in part at the Duke University Shared Materials Instrumentation Facility, a member of the NC Research Triangle Nanotechnology Network, which is supported by the National Science Foundation (award no. ECCS-2025064) as part of the National Nanotechnology Coordinated Infrastructure. K.Y. and X.N. thank support from Hossein (Amir) Salahshoor for providing access to ABAQUS software licensed under SIMULIA. X.N. thanks Heling Wang, Mengdi Han, Nanchao Wang, and Xiaoning Jiang for helpful discussions.

\noindent\textbf{Competing interests:} The authors declare the following competing interests: X.N., K.Y. and S.I. are coinventors on a patent application (U.S. application no. 63/809,023) related to the design, fabrication and applications of the skin-like conformal shape sensor. C.L. and J.Y. hold financial interests in Lumius Imaging, Inc., which had no role in supporting this work. The remaining authors declare no competing interests.
\end{acknowledgments}

\bibliography{ref}

@article{kashyap2020multilayer,
  title={Multilayer fabrication of durable catheter-deployable soft robotic sensor arrays for efficient left atrial mapping},
  author={Kashyap, Varun and Caprio, Alexandre and Doshi, Tejas and Jang, Sun-Joo and Liu, Christopher F and Mosadegh, Bobak and Dunham, Simon},
  journal={Science advances},
  volume={6},
  number={46},
  pages={eabc6800},
  year={2020},
  publisher={American Association for the Advancement of Science}
}

@article{zhang2026flexible,
  title={Flexible skull-conformal phased array for aberration-corrected transcranial focused ultrasound therapy},
  author={Zhang, Jiayi and Hu, Yu and Qin, Haiguo and Pu, Cong and Zhang, Ting and Sun, Junfeng and Peng, Chang},
  journal={Ultrasonics},
  pages={108089},
  year={2026},
  publisher={Elsevier}
}

@inproceedings{boujamaa2008rejection,
  title={Rejection of power supply noise in wheatstone bridges: Application to piezoresistive MEMS},
  author={Boujamaa, El Mehdi and Soulie, Yannick and Mailly, Frederick and Latorre, Laurent and Nouet, Pascal},
  booktitle={2008 Symposium on Design, Test, Integration and Packaging of MEMS/MOEMS},
  pages={96--99},
  year={2008},
  organization={IEEE}
}

@article{yuan2017gelsight,
  title={Gelsight: High-resolution robot tactile sensors for estimating geometry and force},
  author={Yuan, Wenzhen and Dong, Siyuan and Adelson, Edward H},
  journal={Sensors},
  volume={17},
  number={12},
  pages={2762},
  year={2017},
  publisher={MDPI}
}

@article{li2026biomimetic,
  title={Biomimetic multimodal tactile sensing enables human-like robotic perception},
  author={Li, Shoujie and Wu, Tong and Xu, Jianle and Huang, Yan and Zhang, Zongwen and Zhao, Hongfa and Xu, Qinghao and Wang, Zihan and Ye, Linqi and Yang, Yang and others},
  journal={Nature Sensors},
  volume={1},
  number={1},
  pages={52--62},
  year={2026},
  publisher={Nature Publishing Group UK London}
}

@article{xiang2026portable,
  title={A portable and flexible intermediary patch for in vivo magnetic localization},
  author={Xiang, Pingyu and Sun, Danying and Ma, Guoyao and Zhang, Hongye and Xiong, Rong and Wang, Yue and Lu, Haojian and Xu, Tiantian},
  journal={Nature Sensors},
  pages={1--13},
  year={2026},
  publisher={Nature Publishing Group UK London}
}

@article{zhang2022controlling,
  title={Controlling the oxidation and wettability of liquid metal via femtosecond laser for high-resolution flexible electronics},
  author={Zhang, Jingzhou and Zhang, Chengjun and Li, Haoyu and Cheng, Yang and Yang, Qing and Hou, Xun and Chen, Feng},
  journal={Frontiers in Chemistry},
  volume={10},
  pages={965891},
  year={2022},
  publisher={Frontiers Media SA}
}

@article{lu2023robust,
  title={A robust graph-based framework for 3-d shape reconstruction of flexible medical instruments using multi-core fbgs},
  author={Lu, Yiang and Chen, Wei and Li, Bin and Lu, Bo and Zhou, Jianshu and Chen, Zhi and Liu, Yun-Hui},
  journal={IEEE Transactions on Medical Robotics and Bionics},
  volume={5},
  number={3},
  pages={472--485},
  year={2023},
  publisher={IEEE}
}

@article{an2024shape,
  title={Shape reconstruction of soft continuum robots via the fusion of local strains and global poses},
  author={An, Xin and Cui, Yafeng and Dong, Xuguang and Wang, Yixin and Du, Boyuan and Liu, Xin-Jun and Zhao, Huichan},
  journal={Cell Reports Physical Science},
  volume={5},
  number={10},
  year={2024},
  publisher={Elsevier}
}

@article{huang2018three,
  title={Three-dimensional integrated stretchable electronics},
  author={Huang, Zhenlong and Hao, Yifei and Li, Yang and Hu, Hongjie and Wang, Chonghe and Nomoto, Akihiro and Pan, Taisong and Gu, Yue and Chen, Yimu and Zhang, Tianjiao and others},
  journal={Nature electronics},
  volume={1},
  number={8},
  pages={473--480},
  year={2018},
  publisher={Nature Publishing Group UK London}
}

@article{liu2025model,
  title={Model-Based 3D Shape Reconstruction of Soft Robots via Distributed Strain Sensing},
  author={Liu, Liangshu and Huang, Xinghao and Zhang, Xiaoci and Zhang, Baiyu and Xu, Hao and Trivedi, Vedanshee Mihir and Liu, Kenneth and Shaikh, Zoheb and Zhao, Hangbo},
  journal={Soft Robotics},
  year={2025},
  publisher={Mary Ann Liebert, Inc., publishers 140 Huguenot Street, 3rd Floor New~…}
}

@article{lee2020mechano,
  title={Mechano-acoustic sensing of physiological processes and body motions via a soft wireless device placed at the suprasternal notch},
  author={Lee, KunHyuck and Ni, Xiaoyue and Lee, Jong Yoon and Arafa, Hany and Pe, David J and Xu, Shuai and Avila, Raudel and Irie, Masahiro and Lee, Joo Hee and Easterlin, Ryder L and others},
  journal={Nature biomedical engineering},
  volume={4},
  number={2},
  pages={148--158},
  year={2020},
  publisher={Nature Publishing Group UK London}
}

@article{leber2020soft,
  title={Soft and stretchable liquid metal transmission lines as distributed probes of multimodal deformations},
  author={Leber, Andreas and Dong, Chaoqun and Chandran, Rajasundar and Das Gupta, Tapajyoti and Bartolomei, Nicola and Sorin, Fabien},
  journal={Nature Electronics},
  volume={3},
  number={6},
  pages={316--326},
  year={2020},
  publisher={Nature Publishing Group UK London}
}

@article{lilge2022continuum,
  title={Continuum robot state estimation using gaussian process regression on se (3)},
  author={Lilge, Sven and Barfoot, Timothy D and Burgner-Kahrs, Jessica},
  journal={The International Journal of Robotics Research},
  volume={41},
  number={13-14},
  pages={1099--1120},
  year={2022},
  publisher={SAGE Publications Sage UK: London, England}
}

@article{liu2024three,
  title={A three-dimensionally architected electronic skin mimicking human mechanosensation},
  author={Liu, Zhi and Hu, Xiaonan and Bo, Renheng and Yang, Youzhou and Cheng, Xu and Pang, Wenbo and Liu, Qing and Wang, Yuejiao and Wang, Shuheng and Xu, Shiwei and others},
  journal={Science},
  volume={384},
  number={6699},
  pages={987--994},
  year={2024},
  publisher={American Association for the Advancement of Science}
}

@article{xu2025precision,
  title={Precision aerosol-jet micropatterning of liquid metal for high-performance flexible strain sensors},
  author={Xu, Benyan and Yang, Mingyang and Cheng, Wenjun and Li, Xuyin and Xu, Ximei and Li, Wenming and Zhang, Hao and Zhou, Ming},
  journal={Nature Communications},
  volume={16},
  number={1},
  pages={7920},
  year={2025},
  publisher={Nature Publishing Group UK London}
}

@article{huang2023high,
  title={High-stretchability and low-hysteresis strain sensors using origami-inspired 3D mesostructures},
  author={Huang, Xinghao and Liu, Liangshu and Lin, Yung Hsin and Feng, Rui and Shen, Yiyang and Chang, Yuanning and Zhao, Hangbo},
  journal={Science Advances},
  volume={9},
  number={34},
  pages={eadh9799},
  year={2023},
  publisher={American Association for the Advancement of Science}
}

@article{lilge2024state,
  title={State estimation for continuum multi-robot systems on SE (3)},
  author={Lilge, Sven and Barfoot, Timothy D and Burgner-Kahrs, Jessica},
  journal={IEEE Transactions on Robotics},
  year={2024},
  publisher={IEEE}
}

@article{lin2021arei,
  title={Arei: Augmented-reality-assisted touchless teleoperated robot for endoluminal intervention},
  author={Lin, Zecai and Gao, Anzhu and Ai, Xiaojie and Gao, Hongyan and Fu, Yili and Chen, Weidong and Yang, Guang-Zhong},
  journal={IEEE/ASME Transactions on Mechatronics},
  volume={27},
  number={5},
  pages={3144--3154},
  year={2021},
  publisher={IEEE}
}

@article{li2021metasurfaces,
  title={Metasurfaces for bioelectronics and healthcare},
  author={Li, Zhipeng and Tian, Xi and Qiu, Cheng-Wei and Ho, John S},
  journal={Nature Electronics},
  volume={4},
  number={6},
  pages={382--391},
  year={2021},
  publisher={Nature Publishing Group UK London}
}

@article{shah2023stretchable,
  title={Stretchable Shape-Sensing Sheets},
  author={Shah, Dylan and Woodman, Stephanie J and Sanchez-Botero, Lina and Liu, Shanliangzi and Kramer-Bottiglio, Rebecca},
  journal={Advanced Intelligent Systems},
  volume={5},
  number={12},
  pages={2300343},
  year={2023},
  publisher={Wiley Online Library}
}

@article{kim2020heterogeneous,
  title={Heterogeneous sensing in a multifunctional soft sensor for human-robot interfaces},
  author={Kim, Taekyoung and Lee, Sudong and Hong, Taehwa and Shin, Gyowook and Kim, Taehwan and Park, Yong-Lae},
  journal={Science robotics},
  volume={5},
  number={49},
  pages={eabc6878},
  year={2020},
  publisher={American Association for the Advancement of Science}
}

@article{gherlone2018shape,
  title={Shape sensing methods: Review and experimental comparison on a wing-shaped plate},
  author={Gherlone, Marco and Cerracchio, Priscilla and Mattone, Massimiliano},
  journal={Progress in Aerospace Sciences},
  volume={99},
  pages={14--26},
  year={2018},
  publisher={Elsevier}
}

@article{xu2017dual,
  title={Dual-layer orthogonal fiber Bragg grating mesh based soft sensor for 3-dimensional shape sensing},
  author={Xu, Li and Ge, Jia and Patel, Jay H and Fok, Mable P},
  journal={Optics express},
  volume={25},
  number={20},
  pages={24727--24734},
  year={2017},
  publisher={Optica Publishing Group}
}

@article{xu2018stretchable,
  title={Stretchable fiber-Bragg-grating-based sensor},
  author={Xu, Li and Liu, Ning and Ge, Jia and Wang, Xianqiao and Fok, Mable P},
  journal={Optics letters},
  volume={43},
  number={11},
  pages={2503--2506},
  year={2018},
  publisher={Optica Publishing Group}
}

@article{zhu20203d,
  title={3D printed deformable sensors},
  author={Zhu, Zhijie and Park, Hyun Soo and McAlpine, Michael C},
  journal={Science advances},
  volume={6},
  number={25},
  pages={eaba5575},
  year={2020},
  publisher={American Association for the Advancement of Science}
}

@article{majidi2011non,
  title={A non-differential elastomer curvature sensor for softer-than-skin electronics},
  author={Majidi, C and Kramer, R and Wood, RJ},
  journal={Smart materials and structures},
  volume={20},
  number={10},
  pages={105017},
  year={2011},
  publisher={IOP Publishing}
}

@article{white2017multi,
  title={Multi-mode strain and curvature sensors for soft robotic applications},
  author={White, Edward L and Case, Jennifer C and Kramer, Rebecca K},
  journal={Sensors and Actuators A: Physical},
  volume={253},
  pages={188--197},
  year={2017},
  publisher={Elsevier}
}

@article{bai2022dynamically,
  title={A dynamically reprogrammable surface with self-evolving shape morphing},
  author={Bai, Yun and Wang, Heling and Xue, Yeguang and Pan, Yuxin and Kim, Jin-Tae and Ni, Xinchen and Liu, Tzu-Li and Yang, Yiyuan and Han, Mengdi and Huang, Yonggang and others},
  journal={Nature},
  volume={609},
  number={7928},
  pages={701--708},
  year={2022},
  publisher={Nature Publishing Group UK London}
}

@inproceedings{mittendorfer20123d,
  title={3D surface reconstruction for robotic body parts with artificial skins},
  author={Mittendorfer, Philipp and Cheng, Gordon},
  booktitle={2012 IEEE/RSJ International Conference on Intelligent Robots and Systems},
  pages={4505--4510},
  year={2012},
  organization={IEEE}
}

@article{derkevorkian2013strain,
  title={Strain-based deformation shape-estimation algorithm for control and monitoring applications},
  author={Derkevorkian, Armen and Masri, Sami F and Alvarenga, Jessica and Boussalis, Helen and Bakalyar, John and Richards, W Lance},
  journal={AIAA journal},
  volume={51},
  number={9},
  pages={2231--2240},
  year={2013},
  publisher={American Institute of Aeronautics and Astronautics}
}

@article{lun2019real,
  title={Real-time surface shape sensing for soft and flexible structures using fiber Bragg gratings},
  author={Lun, Tian Le Tim and Wang, Kui and Ho, Justin DL and Lee, Kit-Hang and Sze, Kam Yim and Kwok, Ka-Wai},
  journal={IEEE Robotics and Automation Letters},
  volume={4},
  number={2},
  pages={1454--1461},
  year={2019},
  publisher={IEEE}
}

@article{navarro2020model,
  title={A model-based sensor fusion approach for force and shape estimation in soft robotics},
  author={Navarro, Stefan Escaida and Nagels, Steven and Alagi, Hosam and Faller, Lisa-Marie and Goury, Olivier and Morales-Bieze, Thor and Zangl, Hubert and Hein, Bj{\"o}rn and Ramakers, Raf and Deferme, Wim and others},
  journal={IEEE Robotics and Automation Letters},
  volume={5},
  number={4},
  pages={5621--5628},
  year={2020},
  publisher={IEEE}
}

@article{van2018soft,
  title={Soft optoelectronic sensory foams with proprioception},
  author={Van Meerbeek, IM and De Sa, CM and Shepherd, RF},
  journal={Science Robotics},
  volume={3},
  number={24},
  pages={eaau2489},
  year={2018},
  publisher={American Association for the Advancement of Science}
}

@inproceedings{rendl2014flexsense,
  title={FlexSense: a transparent self-sensing deformable surface},
  author={Rendl, Christian and Kim, David and Fanello, Sean and Parzer, Patrick and Rhemann, Christoph and Taylor, Jonathan and Zirkl, Martin and Scheipl, Gregor and Rothl{\"a}nder, Thomas and Haller, Michael and others},
  booktitle={Proceedings of the 27th annual ACM symposium on User interface software and technology},
  pages={129--138},
  year={2014}
}

@article{prutchi1993IDM,
  title={Dynamic contact stress analysis using a compliant sensor array},
  author={Prutchi, David and Arcan, Mircea},
  journal={Measurement},
  volume={11},
  number={3},
  pages={197--210},
  year={1993},
  publisher={Elsevier}
}

@article{kaneko2023printed,
  title={Printed bilayer liquid metal soft sensors for strain and tactile perception in soft robotics},
  author={Kaneko, Takeru and Wang, Yi-Fei and Hori, Mayuka and Sekine, Tomohito and Yoshida, Ayako and Takeda, Yasunori and Kumaki, Daisuke and Tokito, Shizuo},
  journal={Advanced Materials Technologies},
  volume={8},
  number={17},
  pages={2300436},
  year={2023},
  publisher={Wiley Online Library}
}

@article{ma2023shaping,
  title={Shaping a soft future: patterning liquid metals},
  author={Ma, Jinwoo and Krisnadi, Febby and Vong, Man Hou and Kong, Minsik and Awartani, Omar M and Dickey, Michael D},
  journal={Advanced Materials},
  volume={35},
  number={19},
  pages={2205196},
  year={2023},
  publisher={Wiley Online Library}
}

@article{kong2020oxide,
  title={Oxide-mediated mechanisms of gallium foam generation and stabilization during shear mixing in air},
  author={Kong, Wilson and Shah, Najam Ul Hassan and Neumann, Taylor V and Vong, Man Hou and Kotagama, Praveen and Dickey, Michael D and Wang, Robert Y and Rykaczewski, Konrad},
  journal={Soft Matter},
  volume={16},
  number={25},
  pages={5801--5805},
  year={2020},
  publisher={Royal Society of Chemistry}
}

@article{glauser2019deformation,
  title={Deformation capture via soft and stretchable sensor arrays},
  author={Glauser, Oliver and Panozzo, Daniele and Hilliges, Otmar and Sorkine-Hornung, Olga},
  journal={ACM Transactions on Graphics (TOG)},
  volume={38},
  number={2},
  pages={2},
  year={2019},
  publisher={ACM New York, NY, USA}
}

@article{xu20143d,
  title={3D multifunctional integumentary membranes for spatiotemporal cardiac measurements and stimulation across the entire epicardium},
  author={Xu, Lizhi and Gutbrod, Sarah R and Bonifas, Andrew P and Su, Yewang and Sulkin, Matthew S and Lu, Nanshu and Chung, Hyun-Joong and Jang, Kyung-In and Liu, Zhuangjian and Ying, Ming and others},
  journal={Nature communications},
  volume={5},
  number={1},
  pages={3329},
  year={2014},
  publisher={Nature Publishing Group UK London}
}

@article{han2020catheter,
  title={Catheter-integrated soft multilayer electronic arrays for multiplexed sensing and actuation during cardiac surgery},
  author={Han, Mengdi and Chen, Lin and Aras, Kedar and Liang, Cunman and Chen, Xuexian and Zhao, Hangbo and Li, Kan and Faye, Ndeye Rokhaya and Sun, Bohan and Kim, Jae-Hwan and others},
  journal={Nature biomedical engineering},
  volume={4},
  number={10},
  pages={997--1009},
  year={2020},
  publisher={Nature Publishing Group UK London}
}

@article{zhou20233d,
  title={3D Deformation Capture via a Configurable Self-Sensing IMU Sensor Network},
  author={Zhou, Zihong and Chen, Pei and Lu, Yinyu and Cui, Qiang and Pan, Deying and Liu, Yilun and Li, Jiaji and Zhang, Yang and Tao, Ye and Liu, Xuanhui and others},
  journal={Proceedings of the ACM on Interactive, Mobile, Wearable and Ubiquitous Technologies},
  volume={7},
  number={1},
  pages={1--24},
  year={2023},
  publisher={ACM New York, NY, USA}
}

@article{medina2016resistor,
  title={Resistor-based shape sensor for a spatial flexible manifold},
  author={Medina, Oded and Shapiro, Amir and Shvalb, Nir},
  journal={IEEE Sensors Journal},
  volume={17},
  number={1},
  pages={46--50},
  year={2016},
  publisher={IEEE}
}

@article{hu2023stretchable,
  title={Stretchable e-skin and transformer enable high-resolution morphological reconstruction for soft robots},
  author={Hu, Delin and Giorgio-Serchi, Francesco and Zhang, Shiming and Yang, Yunjie},
  journal={Nature Machine Intelligence},
  volume={5},
  number={3},
  pages={261--272},
  year={2023},
  publisher={Nature Publishing Group UK London}
}

@article{scharff2021sensing,
  title={Sensing and reconstruction of 3-D deformation on pneumatic soft robots},
  author={Scharff, Rob BN and Fang, Guoxin and Tian, Yingjun and Wu, Jun and Geraedts, Jo MP and Wang, Charlie CL},
  journal={IEEE/ASME Transactions on Mechatronics},
  volume={26},
  number={4},
  pages={1877--1885},
  year={2021},
  publisher={IEEE}
}

@article{magnet2019,
  title={Soft magnetic skin for continuous deformation sensing},
  author={Hellebrekers, Tess and Kroemer, Oliver and Majidi, Carmel},
  journal={Advanced Intelligent Systems},
  volume={1},
  number={4},
  pages={1900025},
  year={2019},
  publisher={Wiley Online Library}
}

@article{modes2011gaussian,
  title={Gaussian curvature from flat elastica sheets},
  author={Modes, Carl D and Bhattacharya, Kaushik and Warner, Mark},
  journal={Proceedings of the Royal Society A: Mathematical, Physical and Engineering Sciences},
  volume={467},
  number={2128},
  pages={1121--1140},
  year={2011},
  publisher={The Royal Society Publishing}
}

@article{ha2024stretchable,
  title={Stretchable hybrid response pressure sensors},
  author={Ha, Kyoung-Ho and Li, Zhengjie and Kim, Sangjun and Huh, Heeyong and Wang, Zheliang and Shi, Hongyang and Block, Charles and Bhattacharya, Sarnab and Lu, Nanshu},
  journal={Matter},
  volume={7},
  number={5},
  pages={1895--1908},
  year={2024},
  publisher={Elsevier}
}

@article{hao2023online,
  title={An online model-free adaptive tracking controller for cable-driven medical continuum manipulators},
  author={Hao, Jianxiong and Zhang, Kai and Zhang, Zhiqiang and Wang, Shuxin and Shi, Chaoyang},
  journal={IEEE Transactions on Medical Robotics and Bionics},
  year={2023},
  publisher={IEEE}
}

@article{liu2023conformability,
  title={Conformability of flexible sheets on spherical surfaces},
  author={Liu, Siyi and He, Jinlong and Rao, Yifan and Dai, Zhaohe and Ye, Huilin and Tanir, John C and Li, Ying and Lu, Nanshu},
  journal={Science Advances},
  volume={9},
  number={16},
  pages={eadf2709},
  year={2023},
  publisher={American Association for the Advancement of Science}
}

@article{chen2026noise,
  title={A noise-tolerant human--machine interface based on deep learning-enhanced wearable sensors},
  author={Chen, Xiangjun and Lou, Zhiyuan and Gao, Xiaoxiang and Yin, Lu and Qin, Siyu and Lin, Muyang and Zhang, Fangao and Lu, Yi and Ding, Shichao and Liu, Ruixiao and others},
  journal={Nature Sensors},
  volume={1},
  number={1},
  pages={39--51},
  year={2026},
  publisher={Nature Publishing Group UK London}
}

@article{liu2024bioresorbable,
  title={Bioresorbable shape-adaptive structures for ultrasonic monitoring of deep-tissue homeostasis},
  author={Liu, Jiaqi and Liu, Naijia and Xu, Yameng and Wu, Mingzheng and Zhang, Haohui and Wang, Yue and Yan, Ying and Hill, Angela and Song, Ruihao and Xu, Zijie and others},
  journal={Science},
  volume={383},
  number={6687},
  pages={1096--1103},
  year={2024},
  publisher={American Association for the Advancement of Science}
}

@article{zhou2024transcranial,
  title={Transcranial volumetric imaging using a conformal ultrasound patch},
  author={Zhou, Sai and Gao, Xiaoxiang and Park, Geonho and Yang, Xinyi and Qi, Baiyan and Lin, Muyang and Huang, Hao and Bian, Yizhou and Hu, Hongjie and Chen, Xiangjun and others},
  journal={Nature},
  volume={629},
  number={8013},
  pages={810--818},
  year={2024},
  publisher={Nature Publishing Group UK London}
}

@article{hu2023wearable,
  title={A wearable cardiac ultrasound imager},
  author={Hu, Hongjie and Huang, Hao and Li, Mohan and Gao, Xiaoxiang and Yin, Lu and Qi, Ruixiang and Wu, Ray S and Chen, Xiangjun and Ma, Yuxiang and Shi, Keren and others},
  journal={Nature},
  volume={613},
  number={7945},
  pages={667--675},
  year={2023},
  publisher={Nature Publishing Group UK London}
}

@article{zhu20183d,
  title={3D printed functional and biological materials on moving freeform surfaces},
  author={Zhu, Zhijie and Guo, Shuang-Zhuang and Hirdler, Tessa and Eide, Cindy and Fan, Xiaoxiao and Tolar, Jakub and McAlpine, Michael C},
  journal={Advanced Materials},
  volume={30},
  number={23},
  pages={1707495},
  year={2018},
  publisher={Wiley Online Library}
}

@article{al2020improved,
  title={Improved FBG-based shape sensing methods for vascular catheterization treatment},
  author={Al-Ahmad, Omar and Ourak, Mouloud and Van Roosbroeck, Jan and Vlekken, Johan and Vander Poorten, Emmanuel},
  journal={IEEE Robotics and Automation Letters},
  volume={5},
  number={3},
  pages={4687--4694},
  year={2020},
  publisher={IEEE}
}

@article{ha2022shape,
  title={Shape sensing of flexible robots based on deep learning},
  author={Ha, Xuan Thao and Wu, Di and Ourak, Mouloud and Borghesan, Gianni and Dankelman, Jenny and Menciassi, Arianna and Vander Poorten, Emmanuel},
  journal={IEEE Transactions on Robotics},
  volume={39},
  number={2},
  pages={1580--1593},
  year={2022},
  publisher={IEEE}
}

@article{glauser2019interactive,
  title={Interactive hand pose estimation using a stretch-sensing soft glove},
  author={Glauser, Oliver and Wu, Shihao and Panozzo, Daniele and Hilliges, Otmar and Sorkine-Hornung, Olga},
  journal={ACM Transactions on Graphics (ToG)},
  volume={38},
  number={4},
  pages={4},
  year={2019},
  publisher={ACM New York, NY, USA}
}

@article{cluitmans2017vivo,
  title={In vivo validation of electrocardiographic imaging},
  author={Cluitmans, Matthijs JM and Bonizzi, Pietro and Karel, Jo{\"e}l MH and Das, Marco and Kietselaer, Bas LJH and de Jong, Monique MJ and Prinzen, Frits W and Peeters, Ralf LM and Westra, Ronald L and Volders, Paul GA},
  journal={JACC: Clinical Electrophysiology},
  volume={3},
  number={3},
  pages={232--242},
  year={2017},
  publisher={American College of Cardiology Foundation Washington, DC}
}

@article{palieri2020locus,
  title={Locus: A multi-sensor lidar-centric solution for high-precision odometry and 3d mapping in real-time},
  author={Palieri, Matteo and Morrell, Benjamin and Thakur, Abhishek and Ebadi, Kamak and Nash, Jeremy and Chatterjee, Arghya and Kanellakis, Christoforos and Carlone, Luca and Guaragnella, Cataldo and Agha-Mohammadi, Ali-akbar},
  journal={IEEE Robotics and Automation Letters},
  volume={6},
  number={2},
  pages={421--428},
  year={2020},
  publisher={IEEE}
}

@incollection{vlasic2009dynamic,
  title={Dynamic shape capture using multi-view photometric stereo},
  author={Vlasic, Daniel and Peers, Pieter and Baran, Ilya and Debevec, Paul and Popovi{\'c}, Jovan and Rusinkiewicz, Szymon and Matusik, Wojciech},
  booktitle={ACM SIGGRAPH Asia 2009 papers},
  pages={1--11},
  year={2009}
}

@inproceedings{feng2023real,
  title={Real-time element position tracking of flexible array transducer for ultrasound beamforming},
  author={Feng, Z and Hooshangnejad, H and Sforza, D and Vagdargi, P and Bell, MAL and Uneri, A and Ding, K},
  booktitle={Medical Imaging 2023: Ultrasonic Imaging and Tomography},
  volume={12470},
  pages={36--43},
  year={2023},
  organization={SPIE}
}

@article{bai2020stretchable,
  title={Stretchable distributed fiber-optic sensors},
  author={Bai, Hedan and Li, Shuo and Barreiros, Jose and Tu, Yaqi and Pollock, Clifford R and Shepherd, Robert F},
  journal={Science},
  volume={370},
  number={6518},
  pages={848--852},
  year={2020},
  publisher={American Association for the Advancement of Science}
}

@article{wang2022generalizing,
  title={Generalizing to unseen domains: A survey on domain generalization},
  author={Wang, Jindong and Lan, Cuiling and Liu, Chang and Ouyang, Yidong and Qin, Tao and Lu, Wang and Chen, Yiqiang and Zeng, Wenjun and Philip, S Yu},
  journal={IEEE transactions on knowledge and data engineering},
  volume={35},
  number={8},
  pages={8052--8072},
  year={2022},
  publisher={IEEE}
}

@article{yamada2011stretchable,
  title={A stretchable carbon nanotube strain sensor for human-motion detection},
  author={Yamada, Takeo and Hayamizu, Yuhei and Yamamoto, Yuki and Yomogida, Yoshiki and Izadi-Najafabadi, Ali and Futaba, Don N and Hata, Kenji},
  journal={Nature nanotechnology},
  volume={6},
  number={5},
  pages={296--301},
  year={2011},
  publisher={Nature Publishing Group UK London}
}

@book{ochsner2021classical,
  title={Classical beam theories of structural mechanics},
  author={{\"O}chsner, Andreas},
  volume={42},
  year={2021},
  publisher={Springer}
}

@article{darma2020real,
  title={Real-time dynamic imaging method for flexible boundary sensor in wearable electrical impedance tomography},
  author={Darma, PN and Baidillah, MR and Sifuna, MW and Takei, M},
  journal={IEEE sensors journal},
  volume={20},
  number={16},
  pages={9469--9479},
  year={2020},
  publisher={IEEE}
}

@article{li2025flexible,
  title={Flexible intelligent microwave metasurface with shape-guided adaptive programming},
  author={Li, Fan and Pan, Taisong and Li, Weihan and Peng, Zujun and Guo, Dengji and Jia, Xiang and Hu, Taiqi and Wang, Lingxiao and Wang, Wei and Gao, Min and others},
  journal={Nature Communications},
  volume={16},
  number={1},
  pages={3161},
  year={2025},
  publisher={Nature Publishing Group UK London}
}

@article{pan2025shape,
  title={Shape Sensing for Continuum Robots Based on MWCNTs-PDMS Flexible Resistive Strain Sensors},
  author={Pan, Lizhi and Zhang, Tianze and Cheng, Yiding and Ma, Zhikang and Li, Jianmin},
  journal={IEEE Transactions on Medical Robotics and Bionics},
  year={2025},
  publisher={IEEE}
}

\end{document}